\documentstyle[pra,aps,epsf]{revtex}

\newcommand{\beq}{\begin{equation}}
\newcommand{\eeq}{\end{equation}}
\newcommand{\bea}{\begin{eqnarray}}
\newcommand{\eea}{\end{eqnarray}}
\newcommand{\beao}{\begin{eqnarray*}}
\newcommand{\eeao}{\end{eqnarray*}}

\begin{document}
\draft
\twocolumn[\hsize\textwidth\columnwidth\hsize\csname@twocolumnfalse%
\endcsname
\thispagestyle{empty}
\title{
New constraints for non-Newtonian gravity in nanometer range 
from the improved precision measurement
of the Casimir force
}
\author{
M.~Bordag,${}^{1}$
B.~Geyer,${}^{2}$
G.~L.~Klimchitskaya,${}^{3,*}$
 and V.~M.~Mostepanenko${}^{4,*}$
}

\address
{Institute for Theoretical Physics, Leipzig
University,  Augustusplatz 10/11, 04109 Leipzig, 
Germany\\
${}^{1}${Electronic address: Michael.Bordag@itp.uni-leipzig.de}\\
${}^{2}${Electronic address: geyer@rz.uni-leipzig.de}\\
${}^{3}${on leave from North-West Polytechnical 
Institute,St.Petersburg, Russia.\\  
Electronic address:  galina@GK1372.spb.edu}\\
${}^{4}${on leave from A.Friedmann Laboratory
for Theoretical Physics, St.Petersburg, Russia.\\ 
Electronic  address: mostep@fisica.ufpb.br}\\
${}^{*}${Present address: Department of Physics, 
Federal University
of Paraiba,\\ C.P. 5008, CEP 58059-970, Joao Pessoa, Pb-Brazil}
}
%\date{Draft, printed: \today}
\maketitle
\begin{abstract}
We obtain constraints on non-Newtonian gravity following from the
improved precision measurement of the Casimir force by means of
atomic force microscope. The hypothetical force is calculated in
experimental configuration (a sphere above a disk both covered by
two metallic layers). The strengthenings of constraints up to 4 times
comparing the previous experiment and up to 560 times comparing
the Casimir force measurements between dielectrics are obtained in
the interaction range 5.9\,nm$\leq\lambda\leq 115\,$nm. 
Recent speculations
about the presence of some unexplained attractive force in the considered
experiment are shown to be unjustified.
\end{abstract}

]
%%%%%%%%%%%%%%%%%%%%%%%%%%%%%%%%%%%%%%%%%%%%%%%%%%%%%%%%%%%%%%%%% %%%%%
%\narrowtext

Predictions of light and massless elementary particles in unified
gauge theories, supersymmetry, supergravity and string theory have
drawn considerable attention to possible deviations from Newtonian
gravitational law. These deviations can be described by the Yukawa-type
effective potential. Constraints for its parameters (interaction constant
$\alpha$ and interaction range $\lambda$ which is the Compton
wavelength of hypothetical particle) have long been subject for study
(an exhaustive account can be found in \cite{1}). For the range
$10^{-4}\,$m$<\lambda <1\,$m Cavendish- and E\"{o}tvos-type experiments
lead to rather strong constraints on $\alpha$. For larger $\lambda$
the geophysical and satellite measurements are the best to constrain
$\alpha$. No sufficiently strong constraints on $\alpha$ are found,
however, in the range $\lambda< 10^{-4}\,$m.
In this range the existence of Yukawa-type interactions is not excluded
experimentally which are in excess of Newtonian gravitational interaction
by many orders.

For $\lambda< 10^{-4}\,$m the Casimir and van der Waals forces are the
dominant ones acting between electrically neutral macroscopic bodies.
Paper \cite{2} pioneered the application of Casimir and van der Waals
force measurements for constraining the constants of hypothetical
interactions. In succeeding papers [3--7] (see also \cite{8}) the
constraints were obtained both on Yukawa-type and power-type interactions
following from the measurements of Casimir and van der Waals force
between dielectrics performed earlier with an accuracy no better than
10--20\%. These constraints were the best ones in interaction range
$\lambda< 10^{-4}\,$m though they were not so restrictive as the
constraints already known for larger $\lambda$.  

Currently new experiments have been performed on measuring the Casimir
force between metallic surfaces by the use of torsion pendulum \cite{9}
and atomic force microscope \cite{10}. In \cite{9} a 5\% agreement
between theory and experiment was claimed in the whole measurement range
(0.6--6)$\,\mu$m. No experimental evidence was observed for the presence
of finite conductivity, surface roughness or temperature corrections to
the Casimir force. In \cite{10} a 1\% agreement was obtained at the
smallest surface separation, i.e., at a distance 0.12\,$\mu$m. Both, finite
conductivity and surface roughness corrections were taken into account
(temperature corrections are negligible at such separations). There is
a discussion in the literature concerning the ranges of accuracy mentioned
above (see Refs.~\cite{11,12} representing one point of view and
\cite{13,14} representing the other). It should be emphasized particularly,
 however, that the strength of constraints on non-Newtonian
gravity which follow from both experiments does not depend on
estimations of relative error but is determined by the absolute error
of force measurements achieved in both experiments. 

Using this line of reasoning, the constraints following from \cite{9}
were obtained in \cite{15}. In doing so the hypothetical force
together with the corrections to the Casimir force due to finite 
conductivity of
the boundary metal, surface roughness and non-zero temperature were
restricted by the value of $\Delta F=10^{-11}\,$N which is the absolute
error of force measurements in \cite{9}. It was found that the constraints
following from \cite{9} are stronger than the previously known ones up to
a factor of 30 within the range 
$2.2\times 10^{-7}\,$m$\leq\lambda\leq 1.6\times 10^{-4}\,$m (in a more
recent paper \cite{16} a bit different result is obtained; however there
the abovementioned corrections to the Casimir force were not taken into
account).

The constraints on non-Newtonian gravity following from \cite{10} were
obtained in \cite{17}. Here the absolute value of hypothetical force
alone was restricted by the absolute error of force measurements in
\cite{10} $\Delta F=2\times 10^{-12}\,$N (remind that surface roughness and 
finite conductivity corrections to the Casimir force were included into
the measured force). The obtained constraints turned out to be stronger
up to 140 times than the previously known ones within the interaction
range $5.9\,$nm$\leq\lambda\leq 100\,$nm. Notice that some new
possibilities for the search of non-Newtonian gravity in the Casimir
regime are discussed in \cite{18}.

Recently an improved precision measurement of the Casimir force was
performed \cite{19} by the use of atomic force microscope. Owing to the
experimental improvements the absolute error of force measurements was
decreased by a factor of 2 and took the value
$\Delta F=1\times 10^{-12}\,$N. Also the smoother $Al$ coating with
thickness $\Delta_1=250\,$nm was used and much thiner external $Au/Pd$
coating which prevents the oxidation of $Al$ (of the thickness
$\Delta_2=7.9\,$nm). The theoretical corrections to the Casimir force due
to finite conductivity of the metal and surface roughness were calculated
more exactly. This gives the possibility to obtain more strong and
accurate constraints on non-Newtonian gravity in nanometer range.

In the present paper we calculate the gravitational force with account
for non-Newtonian contributions acting in the configuration of experiment
\cite{19}: a polysterene sphere above a sapphire disk both covered by 
thick layers of $Al$ and thin layers of $Au/Pd$. The small distortions covering
the outer surfaces of the test bodies are taken into account whose shape
was investigated by the atomic force microscope. New constraints for the
Yukawa-type hypothetical interaction of nanometer scale coexisting with
Newtonian gravity are obtained. They turned out to be up to 4 times
stronger than the constraints \cite{17} obtained from the previous
experiment \cite{10}. The total strengthening of constraints on
non-Newtonian gravity from the measurements of Casimir force using the
atomic force microscope reaches 560 times within the interval
range $5.9\,$nm$\leq\lambda\leq 115\,$nm.

The potential of gravitational force acting between two atoms with
account of non-Newtonian Yukawa-type contribution can be represented in the
form  
\bea
&&
V(r_{12})=V_N(r_{12})+V_{Yu}(r_{12})
\nonumber\\
&&\phantom{aaaaa}=
-\frac{G M_1 M_2}{r_{12}}\left(1+\alpha_G e^{-r_{12}/\lambda}\right).
\label{1}
\eea
\noindent
Here $M_i=m_p N_i$ are the masses of the atoms, positioned at a distance
$r_{12}$ from each other, $N_i$ are the numbers of nucleons in the atomic
nuclei, $m_p$ is the proton mass. The interaction range
$\lambda=\hbar /(mc)$ has the meaning of the Compton wavelength of a
hypothetical particle of mass $m$. The exchange of this particle between
atoms gives rise to the effective potential $V_{Yu}$ in (\ref{1}).
The new constant $\alpha_G$ characterizes the strength of non-Newtonian
interaction compared to the Newtonian one with a gravitational constant $G$.
In elementary particle physics the effective Yukawa-type potential is
often represented in the form
\begin{equation}
V_{Yu}(r_{12})=-\alpha N_1N_2\hbar c\frac{1}{r_{12}}
e^{-r_{12}/\lambda},
\label{2}
\end{equation}
\noindent
where the multiples $N_i$ are introduced to take off the dependence of
$\alpha$ on the sorts of atoms \cite{20}. Both interaction constants 
$\alpha_G$ and $\alpha$ are connected by the equation
\beq
\alpha_G=\frac{\hbar c}{G m_p^2}\alpha\approx 1.7\times 10^{38}\alpha.
\label{3}
\eeq

Let us calculate the force acting between a disk and a sphere used in the
experiment \cite{19} due to the po\-ten\-tial (\ref{1}). The diameter
$2R=201.7\,\mu$m 
of the sphere is much smaller than the diameter of the disk 
$2L=1\,$cm. Because of this, each atom of the sphere can be considered as if it
would be placed above the center of the disk. 
Let an atom of the sphere with mass $M_1$
be at  height $l\ll L$ above the center of the disk.  
The vertical component of the Newtonian
gravitational force acting between this atom and the disk can be calculated as
\bea
&&
f_{N,z}(l)=\frac{\partial}{\partial l}
\left[GM_1 \rho 2\pi
\int\limits_{0}^{L}r\,dr
\int\limits_{l}^{l+D}
\frac{dz}{\sqrt{r^2+z^2}}\right]
\nonumber\\
&&\phantom{aaa}
\approx -2\pi GM_1\rho D\left[1+\frac{D+2l}{L}\right],
\label{4}\eea
\noindent
where $\rho=4.0\times 10^3\,$kg/m${}^3$ is the sapphire density,
$D=1\,$mm is the thickness of sapphire disk, and only the first-order terms
in $D/L$ and $l/L$ are keeped. 

Integrating Eq.~(\ref{4}) over the volume of the sphere one obtains the
Newtonian gravitational force acting between a sphere and a disk
\beq
F_{N,z}\approx -\frac{8}{3}\pi^2 G\rho\rho^{\prime}DR^3
\left(1+\frac{D}{L}+\frac{2R}{L}\right),
\label{5}
\eeq
\noindent
where $\rho^{\prime}=1.06\times 10^3\,$kg/m${}^3$ is the polysterene
density. Note that this force does not depend on distance $a$ between
the disk and the sphere because of 
$a=(0.1-0.5)\,\mu$m$\ll R$.

Substituting the values of parameters given above into (\ref{5}) we
arrive to the value $F_{N,z}\approx 7.6\times 10^{-18}\,$N which falls
far short of the absolute error of force mesurements
$\Delta F=1\times 10^{-12}\,$N in \cite{19}. The value of Newtonian
gravitational force between the test bodies remains  nearly unchanged
when taking into account 
the contributions of $Al$ and $Au/Pd$ layers on the sphere and the disk. 
The corresponding result can be simply obtained by the
combination of several expressions of type (\ref{5}). The additions to
the force due to layers are suppressed by the small multiples
$\Delta_i/D$ and $\Delta_i/R$. That is why the Newtonian gravitational
force is negligible in experimental configuration of \cite{19}
(note that for configuration of two plane parallel plates gravitational
force can play more important role \cite{18}).

Now we consider a non-Newtonian force acting between a disk and a sphere
due to the potential $V_{Yu}$ from (\ref{1}), (\ref{2}). It can be calculated
most simply using the same procedure as in a Newtonian case. For a
homogeneous sphere of density $\rho^{\prime}$ and a disk of density $\rho$
the result is
\bea
&&
F_{Yu,z}(a)=-4\pi^2 G\alpha_G \rho\rho^{\prime}
\lambda^3
e^{-a/\lambda}
\left[\vphantom{e^{-2R/\lambda}}
R-\lambda\right.
\label{6}\\
&&\phantom{aaaaaaaa}
+\left.
(R+\lambda)\,e^{-2R/\lambda}\right].
\nonumber
\eea

In the case of Yukawa-type long-range interaction of unknown strength we
should carefully take account the covering layers of thickness
$\Delta_1$ ($Al$) and $\Delta_2$ ($Au/Pd$). Combining 25 contributions of the
form of (\ref{6}) with regard to $a,\lambda\ll R$ one obtaines \cite{17}

\bea
&&F_{Yu}(a)=-4\pi^2G \alpha_G 
\lambda^3
e^{-\frac{a}{\lambda}}\,R
\left[\rho_2-(\rho_2-\rho_1)
e^{-\frac{\Delta_2}{\lambda}}
\right.
\nonumber\\
&&\phantom{aaaaa}\left.
-(\rho_1-\rho)
e^{-\frac{\Delta_2+\Delta_1}{\lambda}}\right]
\left[\rho_2 -
(\rho_2-\rho_1)
e^{-\frac{\Delta_2}{\lambda}}\right.
\nonumber \\
&&\phantom{aaaaa}
-\left.
(\rho_1-\rho^{\prime})
e^{-\frac{\Delta_2+\Delta_1}{\lambda}}
\right].
\label{7}
\eea
\noindent
Here $\rho_1=2.7\times 10^3\,$kg/m${}^3$ is the density of $Al$ and
 $\rho_2=16.2\times 10^3\,$kg/m${}^3$ is the density of 60\%$Au$/40\%$Pd$.

As it was shown in \cite{17}, the surface distortions can significantly
influence the value of hypothetical force in the nanometer range. 
In \cite{19}  smoother metal coatings than in \cite{10} were used.
The roughness of the metal surface was measured with the atomic force
microscope. The major distortions both on the disk and on the sphere
can be modeled by parallelepipeds of two heights
$h_1=14\,$nm (covering the fraction $v_1=0.05$ of the surface) and
$h_2=7\,$nm (which cover the fraction 
$v_1=0.11$  of the surface). The surface 
between these distortions is covered by a stochastic roughness of
height $h_0=2\,$nm ($v_0=0.84$ fraction of the surface). It consists
of small crystals which form a homogeneous background of the averaged
height $h_0/2$.

The height $H$ relative to which the mean value of the roughness
function is zero can be found from the equation
\beq
(h_1-H)v_1+(h_2-H)v_2-\left(H-\frac{h_0}{2}\right)v_0=0.
\label{8}
\eeq
\noindent
From (\ref{8}) one obtains $H=2.31\,$nm. The roughness functions of the
disk and of the sphere can be defined by
\bea
&&
z_1^{(s)}=Af_1(x_1,y_1),
\label{9}\\
&&
z_2^{(s)}=a+R-\sqrt{R^2-r^2}+Af_2(x_2,y_2),
\nonumber
\eea
\noindent
where the amplitude $A$ is chosen in such a way that
$\max|f_i(x_i,y_i)|=1$. The value of the amplitude defined relatively to
zero distortion level is
\beq
A=h_1-H=11.69\,\mbox{nm}.
\label{10}
\eeq
\noindent
In terms of the parameters
\beq
\beta_1=\frac{h_2-H}{A}\approx 0.4012,
\qquad
\beta_2=\frac{H-h_0/2}{A}\approx 0.1121
\label{11}
\eeq
\noindent
the distortion functions from Eq.(\ref{9}) can be represented as
\beq
f_1(x_1,y_1)=\left\{
\begin{array}{rcl}
1, && (x_1,y_1)\in \Sigma_{s_1},\\   
\beta_1, && (x_1,y_1)\in \Sigma_{s_2},\\
-\beta_2 && (x_1,y_1)\in \Sigma_{s_0},
\end{array}\right.
\label{12}
\eeq
\[
f_2(x_2,y_2)=\left\{
\begin{array}{rcl}
-1, && (x_2,y_2)\in \tilde{\Sigma}_{s_1},\\   
-\beta_1, && (x_2,y_2)\in \tilde{\Sigma}_{s_2},\\
\beta_2 && (x_2,y_2)\in \tilde{\Sigma}_{s_0},
\end{array}\right.
\nonumber
\]
\noindent
where $\Sigma_{s_i}$ ($\tilde{\Sigma}_{s_i}$) are the regions of the
disk (sphere) occupied by the distortions of different kinds.

To find the Yukawa force with account of roughness we calculate
the values of force (\ref{7}) for six different distances which occur
between the distorted surfaces, multiply them by the appropriate
probabilities, and make a summation
\bea
&&
F_{Yu}^R(a)=
\sum\limits_{i=1}^{6}w_iF_{Yu}(a_i)\equiv
v_1^2F_{Yu}(a-2A)
\nonumber\\
&&\phantom{a}
+2v_1v_2F_{Yu}\left(a-A(1+\beta_1)\right)
\label{13}\\
&&\phantom{a}
+
2v_2v_0F_{Yu}\left(a-A(\beta_1-\beta_2)\right)+
v_0^2F_{Yu}(a+2A\beta_2)
\nonumber\\
&&\phantom{a}
+v_2^2F_{Yu}(a-2A\beta_1)
+2v_1v_0F_{Yu}\left(a-A(1-\beta_2)\right).
\nonumber
\eea
\noindent
Note that exactly the same result can be obtained 
for the Casimir force by the use of
perturbation expansion up to fourth order in powers of relative
distortion amplitude \cite{14,21}.

Let us consider now the constraints on non-Newtonian gravity following
from \cite{19}. According to the results of \cite{19} no hypothetical
force was observed. This means that the hypothetical forces in question
are constrained by the inequality
\beq
|F_{Yu}^R(a)|\leq\Delta F=1\times 10^{-12}\,\mbox{N},
\label{14}
\eeq
\noindent
where $F_{Yu}^R(a)$ is calculated in (\ref{7}), (\ref{13}).

The strongest constraints on $\alpha_G$ follow from Eq.~(\ref{14}) for
the smallest possible value of $a$. There is $a_{\min}=100\,$nm in the Casimir
force measurement of \cite{19}. This distance is between $Al$ layers
because the $Au/Pd$ layers of $\Delta_2=7.9\,$nm thickness were shown
to be transparent for the essential frequences of order $c/a$.
Considering the Yukawa-type hypothetical interaction
this means that $a_{\min}^{Yu}=100\,\mbox{nm}-2\Delta_2=84.2\,$nm.
Substituting this value into (\ref{14}) with account of (\ref{7}),
(\ref{13}) one obtains constraints on $\alpha_G$ following from the
experiment \cite{19} for different $\lambda$. The computational results
are presented in Fig.~1 by the curve 4. In the same figure the curve 1
indicates the known constraints \cite{4,6,8} on the Yukawa-type
interaction following from the old measurements of the Casimir force
between dielectrics \cite{22,23}. The curve 2 shows the known
constraints \cite{6,7} following from the measurements of van der Waals
force \cite{24}. By the curve 3 the constraints are shown \cite{17}
following from the first Casimir force measurement by means of atomic
force microscope \cite{10}. The region below each curve in the
($\lambda,\alpha_G$) plane is permitted and above the curve is prohibited.
For convenience in addition to the vertical axis $\alpha_G$ the vertical axis
$\alpha$ is introduced to illustrate the relation between the obtained
constraints in terms of the constants $\alpha_G$ and $\alpha$ connected by
Eq.~(\ref{3}).

As is seen from Fig.~1, the improved precision measurement of the
Casimir force gives the possibility to strengthen the constraints shown
by the curve 3 up to 4 times within a range
5.9\,nm$\leq\lambda\leq 115\,$nm. The largest strengthening takes place for
$\lambda=(10-15)\,$nm. Comparing the previously known constraints in
nanometer range (curves 1 and 2 in Fig.~1) the strengthening up to 560
times was achieved by the Casimir force measurement using the atomic force
microscope. Also the improved measurement gave the possibility to widen
the range where the stronger constraints are obtained till 115\,nm
(instead of 100\,nm in \cite{17}).
There is a gap till the value $\lambda=220\,$nm starting from which the
stronger constraints follow from the experiment \cite{9}. In this gap
the old constraints obtained from the Casimir force measurements between
dielectrics \cite{6} are still the best ones. 
Note that with the use of smoother metal
coatings the contribution of surface roughness to the Yukawa force is
decreased. For example, for $\lambda=6\,$nm the point of the curve 4
in Fig.~1 calculated with account of roughness is
log$\alpha=-15.17$ and without account of roughness log$\alpha=-15.11$.
For $\lambda=20\,$nm one obtaines log$\alpha=-20.41$ with account of
roughness and log$\alpha=-20.44$ for the absolutely smooth test bodies.
For larger $\lambda$ the influence of roughness is even smaller.

It is significant
that in the recent paper \cite{25} the attempt was undertaken to
calculate the Casimir force in experimental configuration of \cite{19}
with account of outer $Au/Pd$ layers. The result was obtained that
the maximum possible theoretical value of the force is significantly 
smaller than the measured ones. According to the speculations of
\cite{25} the observed discrepancy is explained by a new Yukawa force
mediated by a light scalar boson. This conclusion is, however,
incorrect for the following reason.
We remind that in \cite{19} as space separations the distances between
$Al$ layers have been taken.
To take into
account the $Au/Pd$ layers the authors of \cite{25} changed the experimental 
data of \cite{19} ``by shifting all the points to larger separations on
$2h=16\,$nm'' (where $h=8\,$nm is the layer thickness in \cite{19}).
Instead of this the shift to smaller separations by 16\,nm should be done 
to get the actual distance between the outer $Au/Pd$ layers. If the correct 
shift is done then the theoretical values of the force, including the
effect of covering layers, are in accordance with the experimental values 
in the limits of absolute error of force measurements (\ref{14}).
Hence the conclusion of \cite{25} about the probable influence of new
hypothetical attraction based on experiment \cite{19} is unjustified.

To conclude  stronger constraints on the non-Newtonian gravity were
obtained from the improved precision measurement of the Casimir force using
the atomic force microscope. Their strength exceeds up to 560 times the 
previously known constraints in nanometer range
5.9\,nm$\leq\lambda\leq 115\,$nm. Further strengthening of constraints on
non-Newtonian gravity from the Casimir effect is expected in the future.

\section*{ACKNOWLEDGMENTS}  

The authors are grateful to U.~Mohideen for 
several helpful discussions. 
G.L.K.\ and V.M.M.\ are indebted 
to Center of Theoretical Sciences and  
Institute of Theoretical Physics of Leipzig University,
where this work was performed, for kind hospitality. G.L.K.
\ was supported by Graduate College on Quantum Field Theory at Leipzig
University.  V.M.M.\ was
supported by Saxonian Ministry for Science and Fine Arts.

%%%%%%%%%%%%%%%%%%%%%%%%%%%%%%%%%%%%%%%%%%%%%%%%%%%%%%%%%%%%%%%%%%%%%%%%%%%%%

\begin{figure}[h]
\epsfxsize=10cm\centerline{\epsffile{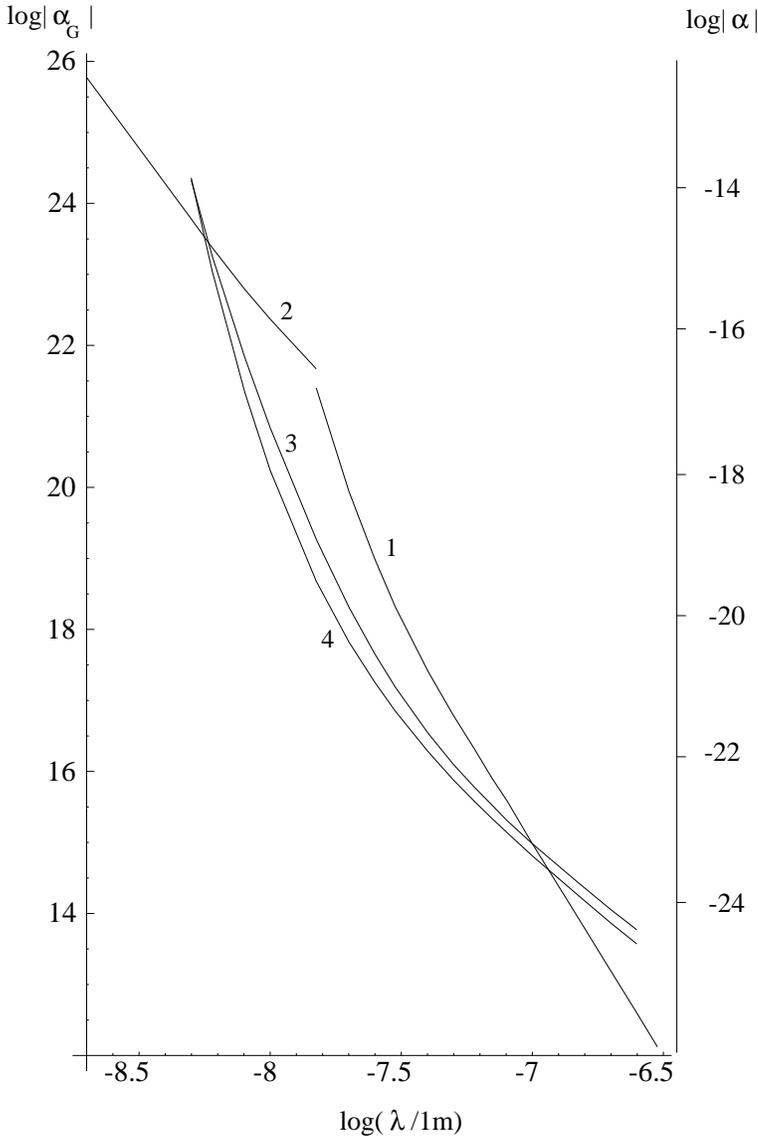} }
\caption{
Constraints for the constants of
Yukawa-type hypothetical interactions.
Curves 1 and 2 follow respectively
from the Casimir and van der Waals force measurements  between dielectrics
[6,22--24].
Curve 3 follows [17] from the first  Casimir force 
 measurement  using
an atomic force microscope [10].
Curve 4 is obtained in this paper from the improved precision
measurements of the Casimir force [19]. The regions above the curves are
prohibited, and below the curves are permitted.
}
\end{figure}

\begin{thebibliography}{99}
\bibitem{1}
E.~Fischbach and C.~L.~Talmadge,
{\it The Search for Non-Newtonian Gravity}
(Springer-Verlag, New York, 1998).
\bibitem{2} 
V.~A.~Kuz'min, I.~I.~Tkachev, and 
M.~E.~Shaposhnikov, JETP Letters
  (USA) {\bf 36}, 59 (1982).
\bibitem{3} 
V.~M.~Mostepanenko and I.~Yu.~Sokolov, Phys. Lett. A {\bf 125}, 405
(1987).
\bibitem{4} 
V.~M.~Mostepanenko and I.~Yu.~Sokolov, Phys. Lett. A {\bf 132}, 313
(1988).
\bibitem{5} 
Yu.~N.~Moiseev, V.~M.~Mostepanenko, V.~I.~Panov,
 and I.~Yu.~Sokolov, Sov. Phys. --- Dokl. (USA) {\bf 34}, 147
(1989).
\bibitem{6} 
V.~M.~Mostepanenko and I.~Yu.~Sokolov, Phys. Rev. D {\bf 47}, 2882
(1993).
\bibitem{7} 
M.~Bordag, V.~M.~Mostepanenko, and I.~Yu.~Sokolov, 
Phys. Lett. 
A {\bf 187}, 35 (1994).
\bibitem{8} 
V.~M.~Mostepanenko and N.~N.~Trunov, 
{\it The Casimir Effect and 
Its Applications} (Clarendon Press, Oxford, 1997).
\bibitem{9} 
S.~K.~Lamoreaux, Phys. Rev. Lett. {\bf 78}, 5 (1997); {\bf 81}, 5475(E) (1998).
\bibitem{10} 
U.~Mohideen and A.~Roy, 
Phys. Rev. Lett. {\bf 81}, 4549 (1998).
\bibitem{11} 
S.~K.~Lamoreaux, Phys. Rev. A {\bf 59}, R3149 (1999).
\bibitem{12} 
S.~K.~Lamoreaux, Phys. Rev. Lett. {\bf 83}, 3340 (1999).
\bibitem{13} 
U.~Mohideen and A.~Roy, 
Phys. Rev. Lett. {\bf 83}, 3341 (1999).
\bibitem{14} 
G.~L.~Klimchitskaya, A.~Roy,  U.~Mohideen, and 
V.\ M.\ Mostepanenko, Phys. Rev. A {\bf 60}, 3487 (1999). 
\bibitem{15} 
M.~Bordag, B.~Geyer,  G.~L.~Klimchitskaya, 
and V.\ M.\ Mostepanenko,
   Phys. Rev. D {\bf 58}, 075003  (1998).
\bibitem{16}
J.~C.~Long, H.~W.~Chan, and J.~C.~Price,
Nucl. Phys. B {\bf 539}, 23 (1999).
\bibitem{17} 
M.~Bordag, B.~Geyer,  G.~L.~Klimchitskaya, 
and V.\ M.\ Mostepanenko,
   Phys. Rev. D {\bf 60}, 055004  (1999).
\bibitem{18} 
 D.~E.~Krause and E.~Fischbach,
To appear in {\it Testing General Relativity in Space:
Gyroscopes, Clocks, and Interferometers}, edited by
C.~L\"{a}mmerzahl, C.W.F.~Everitt, F.W.~Hehl (Springer-Verlag, 2000);
hep-ph/9912276.
\bibitem{19} 
A.~Roy, C.-Y.~Lin, and U.~Mohideen, 
Phys. Rev. D {\bf 60}, R111101 (1999).
\bibitem{20}
G.~Feinberg and J.~Sucher, 
Phys. Rev. D {\bf 20}, 1717 (1979).
\bibitem{21}
V.~B.~Bezerra, G.~L.~Klimchitskaya, and C.~Romero,
Phys. Rev. A {\bf 61}, 022115 (2000).
\bibitem{22} 
B.~V.~Derjaguin, I.~I.~Abrikosova, 
and E.~M.~Lifshitz, Quart. Rev. Chem. Soc. 
{\bf 10}, 295 (1956).
\bibitem{23} 
S.~Hunklinger, H.~Geisselmann, and 
W.~Arnold, Rev. Sci. Instr. 
{\bf 43}, 584 (1972).
\bibitem{24} 
Y.~N.~Israelachvili and D.~Tabor,  
Proc. Roy. Soc. Lond. 
A {\bf 331}, 19 (1972).
\bibitem{25}
V.~B.~Svetovoy and M.~V.~Lokhanin, quant-ph/0001010.
\end{thebibliography}
\end{document}